\documentclass[epj]{svjour}
\usepackage{graphicx}
\usepackage{epstopdf}
\usepackage{amssymb}
\usepackage{bm}
\usepackage{amsmath}
\usepackage{nicefrac}
\usepackage{siunitx}
\usepackage{tikz}
\usepackage{lipsum}
\usepackage{enumitem}
\begin{document}
\title{
Ultraviolet suppression and nonlocality in optical model potentials for
nucleon-nucleus scattering}

\author{
  H. F. Arellano\inst{1} 
  \and 
  G. Blanchon\inst{2,3}% etc
% \thanks is optional - remove next line if not needed
%\thanks{\emph{Present address:} Insert the address here if needed}%
}                     % Do not remove
%
%\offprints{}          % Insert a name or remove this line
%

\institute{
  Department of Physics - FCFM, University of Chile,
  Av. Blanco Encalada 2008, Santiago, Chile
 \and
  CEA,DAM,DIF, F-91297 Arpajon, France
  \and
  Universit\'e Paris-Saclay, CEA, Laboratoire Mati\`ere en 
  Conditions Extr\^emes,Bruy\`eres-le-Ch\^atel, France
          }
\date{Received: date / Revised version: date}
% The correct dates will be entered by Springer
%
\abstract{
  We investigate the role of high momentum components of
  optical model potentials for nucleon-nucleus scattering
  and its incidence on their nonlocal structure in coordinate space.
  The study covers closed-shell nuclei with mass number in the range 
  $4\!\leq\! A\!\leq\! 208$, for nucleon 
  energies from tens of MeV up to 1~GeV.
  To this purpose microscopic optical potentials were calculated
  using density-dependent off-shell $g$ matrices in 
  Brueckner-Hartree-Fock approximation and based on Argonne $v_{18}$
  as well as chiral 2\emph{N} force up to 
next-to-next-to-next-to-leading order.
  We confirm that the gradual suppression of high-momentum contributions 
  of the optical potential results in quite different 
  coordinate-space counterparts, all of them accounting for 
  the same scattering observables.
  We infer a minimum cutoff momentum $Q$, a function of the target mass
  number and energy of the process, that filters out irrelevant 
  ultraviolet components of the potential.
  We find that when ultraviolet suppression is applied to Perey-Buck
  nonlocal potential or local Woods-Saxon potentials, they 
  result with similar nonlocal structure to those obtained from
  microscopic models in momentum space.
  We examine the transversal nonlocality, quantity that makes 
  comparable the intrinsic nonlocality of any potential
  regardless of its representation.
% Findings of this work may help to bridge the gap between momentum-
% and coordinate-space developments for optical model potentials
% for nucleon-nucleus scattering.
  We conclude that meaningful comparisons of nonlocal features of 
  alternative potential models require the suppression of their ultraviolet
  components.
 % end of PACS codes
} %end of abstract
\authorrunning{Arellano-Blanchon}
\titlerunning{Ultraviolet suppression and nonlocality...}
\maketitle
 \sloppy
\section{Introduction}
\label{intro}
It is a broadly accepted fact that optical model potentials
for nucleon-nucleus (\textit{NA}) scattering are energy-dependent, 
complex and nonlocal operators. 
Their nonlocality arises from the fermionic nature of the interacting 
nucleons in conjunction with intrinsic nonlocalities of 
nucleon-nucleon (\textit{NN}) interactions.
By locality it is alluded to a particular structure of the
interaction in coordinate space, being diagonal in the pre-
and post-collision relative coordinates. 
An early departure from this construction was introduced by Perey and 
Buck (PB) in the early 60s~\cite{Perey1962}, 
with the inclusion of a phenomenological finite-width
Gaussian form factor in the central part of the potential. 
The width of the Gaussian is customary used to quantify the degree
of nonlocality and is still broadly 
used~\cite{Atkinson2020,Pruitt2020,Perez2019,Tian2018}.

In a recent work~\cite{Arellano2018}
we have investigated the nonlocal structure of microscopic 
folding optical-model potentials calculated in momentum space.
The study focuses on proton scattering off $^{40}$Ca at energies
from 30~MeV up to 1~GeV.
An important result of that investigation is that scattering
observables and associated wavefunctions remain invariant under
the suppression of momentum components of the potential above
some cutoff momentum $\Lambda$. 
Interestingly, the implied potentials in coordinate space
exhibit quite different nonlocal structure.
In this work we elaborate further those finding by considering
targets over the mass range $4\leq A\leq 208$.
We find that the suppression of high momentum
components of any potential leads to equivalent ones with
similar shapes in coordinate space.

Historically, the construction and calculation of optical model 
potentials has adopted routes spanning from pure
phenomenological models to strictly microscopic ones. 
Additionally, they can either be developed in coordinate or
momentum representations.
Furthermore,
within the coordinate-space class, 
they can also be subdivided into local and nonlocal ones.
On each of these approaches there are special features
of the potential which are often scrutinized such as
depth, radii, diffuseness, nonlocality, and off-shellness,
to mention some of the most common.
Comparisons among these models can be made only at the end point,
after solving Schr\"odinger equation,
assessing their scattering amplitudes and 
level of agreement with scattering measurements.

Woods-Saxon potentials constitute a classic example of phenomenological
local potential in coordinate space, where some parameters
are adjusted in order to reproduce scattering data. In this work
we pay attention to the global optical model by
Koning-Delaroche~\cite{Koning2003},
developed for nucleon energies of up to 200~MeV.
The inclusion of nonlocality introduced by Perey-Buck
folds a nonlocal form factor in the central part of the potential. 
The nonlocality is controlled through the width $\beta$ 
of a Gaussian form factor, typically of the order of 0.8~fm.
A recent parametrization of PB model has been introduced by 
Tian-Pang-Na~\cite{Tian2015} (TPM), 
enabling the study of
proton scattering at nucleon beam energies of up to 30~MeV.

Microscopic optical potentials have the interesting feature to
provide a link between 
the bare \textit{NN} interaction and
the $(A\!+\!1)$-body system.
Pioneering work along this line was introduced by Brieva and 
Rook~\cite{Brieva1977}, with the first microscopic folding approach
for \textit{NA} scattering.
Simultaneously Jeukenne, Lejeune and Mahaux~\cite{Jeukenne1976},
introduce the local density approximation for the optical model
potential.
Here, at each coordinate $r$ of the projectile in the nucleus,
the on-shell mass operator from infinite nuclear matter
--evaluated at the density of the target at coordinate $r$--
is mapped to the local potential.
The energy at which the mass operator is evaluated corresponds to
that of the beam.
%Besides the fact that the resulting optical potential is local,
%its spin-orbit part becomes undefined.
%This limitation is fixed with the introduction of a phenomenological 
%spin-orbit coupling /cite{REF}.

Based on these developments,
local \textit{NN} effective interactions
were introduced by von Geramb~\cite{Geramb1983} and
subsequently refined by Amos and collaborators~\cite{Amos2000},
to be used in the calculation of microscopic nonlocal optical
potentials in coordinate space. 
The resulting strengths of Yukawa form factors
of Hamburg and Melbourne \textit{NN} effective interactions 
have been embedded
in the DWBA98 computational code developed by Jacques 
Raynal~\cite{Raynal1998}, where the nonlocal part of the potential
arises from the exchange term of the interaction.
Applications of this approach can be made from few tens of MeV up to
about 300~MeV.

Momentum-space 
microscopic folding approaches for \textit{NA} 
scattering have been extensively investigated since the 
mid 80s~\cite{Picklesimer1984}.
Subsequent developments led to the so called
full-folding approach for the optical model 
potential~\cite{Ray1992,Arellano1989,Arellano1990,Elster1990,Crespo1990,Vorabbi2016}.
Here a convolution takes place between an \textit{NN} 
effective interaction 
(off-shell $g$- or free $t$-matrix, 
depending on the energy of application)
and the ground-state nonlocal one-body mixed density of the target.
Further developments within momentum-space folding approaches
include the account for nuclear medium effects, as
governed by genuine off-shell $g$ 
matrices~\cite{Arellano1995,Arellano2002,Aguayo2008}.
An appealing feature of these approaches is that the
nonlocality of the \textit{NA} potential is naturally accounted for,
although such features remain undisclosed.
In this work we present a means to compare them 
regardless their representation or local/nonlocal features.
Along the process we are able to separate those Fourier components
of the potential which are inherent to the \textit{NA} scattering
process from those that turn out irrelevant in the context 
of Schr\"odinger's wave equation.

This work is organized as follows.
In sec.~\ref{sec:framework} we layout the theoretical framework
and present the optical model approach to be considered as benchmark.
In sec.~\ref{sec:findings} we study the effect of high momentum components
of the optical potential, set the threshold momentum above which 
scattering observables are invariant, assess the nonlocality of
momentum and coordinate (local and nonlocal) potentials after
ultraviolet suppression, and propose a means to assess their
nonlocality.
In sec.~\ref{sec:summary} we present a summary of the major findings
of this work and the main conclusions.
We also include an Appendix for explicit formulas used for
Gaussian multipoles.

\section{Framework}
\label{sec:framework}
In the context of nucleon scattering off nuclei we express
the optical model potential in momentum space as the
sum of central and spin-orbit contributions 
\begin{equation}
\label{omp}
\tilde U({\boldsymbol k}',{\boldsymbol k};E) =
\tilde U_{c}({\boldsymbol k}',{\boldsymbol k};E) +
i {\bm \sigma}\cdot \hat n\;
\tilde U_{so}({\boldsymbol k}',{\boldsymbol k};E)\;,
\end{equation}
where ${\textstyle{\frac12}}{\bm\sigma}$ corresponds to the spin of the 
projectile 
and $\hat n$ is a unit vector perpendicular to the scattering plane,
with
${\bm k'}\times{\bm k} = \hat n\,|{\bm k'}\times{\bm k}|$. 
Operator $\hat U$ in Eq.~\eqref{omp} is also denoted as
$\tilde\Sigma({\boldsymbol k}',{\boldsymbol k};E)$ 
or $\tilde M({\boldsymbol k}',{\boldsymbol k};E)$ by other authors
\cite{Bell1959,Rotureau2017}.

Actual calculations of the potential in momentum space 
are performed over a finite mesh of relative momenta $k$ up to 
some maximum value ranging from 10 up to 20~fm$^{-1}$,
depending on the beam energy and target.
Additionally, an angular mesh $\hat k\cdot\hat k'$ is 
designed for reliable partial wave expansion.
Once the central and spin-orbit components of
$\tilde U({\boldsymbol k}',{\boldsymbol k};E)$ are obtained,
the corresponding partial wave components $\tilde U_{jl}(k',k)$ 
can be calculated, with $j$ and $l$ the total and orbital
angular momenta, respectively. 
Detailed expressions in the context of this construction can be
found in Ref.~\cite{Arellano2021}.

\subsection{Ultraviolet suppression}
\label{sec:ultraviolet}
Even though the optical potential is calculated
in momentum space,
we carry out the calculation of scattering waves and observables 
in coordinate space~\cite{Arellano2021}. 
Thus, for a given $\hat U({\bm k}',{\bm k})$ we perform double 
Fourier Transforms (FT), which in the case of the central component 
of the potential takes the form
\begin{equation}
\label{kk2rr}
  U_{l}(r',r) = \frac{2}{\pi}
  \int_0^\infty k'^2dk' 
  \int_0^\infty k^2dk\,
  j_l(k'r') \tilde U_{l}(k',k) j_l(kr)\;.
\end{equation}
In general this double integral results in a non-diagonal
(nonlocal) function in $r,r'$ coordinates
(we omit subscript $c$ for simplicity).
Similar expressions hold for the spin-orbit term.
Evaluations of the above integrals are performed up to some 
upper momenta chosen to ensure convergence of 
scattering observables. Symbolically,
\begin{equation}
  \label{xk1k2}
  U_{l}(r',r) = 
  \frac{2}{\pi}
  \int_{0}^{\infty} \!dk'\!
  \int_{0}^{\infty} \!dk \cdots \to
  \frac{2}{\pi}
  \int_{0}^{\Lambda} \!dk'\!
  \int_{0}^{\Lambda} \!dk \cdots \;,
\end{equation}
representing an ultraviolet cutoff of the interaction.
In practice, the only condition imposed to $\Lambda$ is
that scattering observables remain invariant under its variations.
With this guiding criterion we look for the minimum $\Lambda$ 
as a function of the target mass $A$ and the beam energy.

For the ultraviolet cutoff we use the hyperbolic 
regulator $f_\Lambda(k)$ defined as
\begin{equation}
\label{cutoff}
f_\Lambda(k)=
\frac12\left [1-\tanh\left (\frac{k-\Lambda}{\delta}\right )\right ]\;,
\end{equation}
which in the limit $\delta\to 0$,
becomes the Heaviside step function $\Theta(\Lambda - k)$.
In this study we use $\delta=0.2$~fm$^{-1}$, as width of the cutoff.
In what follows we focus on the implications of this cutoff,
namely
\begin{equation}
\label{ulambda}
\tilde U(k',k) \to
\tilde U_\Lambda(k',k) =
f_\Lambda(k') \tilde U(k',k) f_\Lambda(k)\;.
\end{equation} 

\subsection{Optical-model and scattering calculations}
\label{sec:calculations}

To narrow margins of arbitrariness in the \textit{NA} coupling
we consider a single
microscopic approach applicable to a wide energy range.
To this purpose we follow Ref.~\cite{Aguayo2008}
for momentum-space constructions,
where an \textit{in-medium} \textit{NN} effective interaction is folded
with the target full mixed density.
The nonlocal density-dependent effective interaction is 
taken as actual off-shell $g$ matrix, 
solution of the Brueckner-Bethe-Goldstone equation
in the Brueckner-Hartree-Fock approximation.
In absence of medium effects the $g$ matrix becomes
the scattering $t$ matrix, resulting in the impulse approximation 
for the optical model potential in multiple-scattering 
expansion 
\cite{Arellano1990,Elster1990,Crespo1990,Vorabbi2016,Weppner1998}.
The momentum-space folding approach we follow constitutes a genuine
parameter-free description of nucleon scattering off nuclei 
at energies ranging from few tens of MeV up to 1~GeV
\cite{Arellano1995,Arellano2002,Aguayo2008}.

Nuclear-matter $g$ matrices are based on
the traditional Argonne 
$v_{18}$ \cite{Wiringa1995} (AV18) bare potential
fitted to \emph{NN} phase-shift data at beam energies
below pion production threshold, together with static properties 
of the deuteron. 
Additionally, 
we include results based on chiral effective-field-theory 
interaction. 
In this case the bare interaction is constructed with nucleons and 
pions as degrees of freedom, with the two-nucleon part (2\emph{N}) 
fit to \emph{NN} data. 
We use the chiral 2\emph{N} force up to 
next-to-next-to-next-to-leading order (N3LO) given 
by Entem and Machleidt \cite{Entem2003}.
For each of these interactions we have calculated the
corresponding infinite nuclear matter self-consistent single-particle
fields following Refs.  \cite{Arellano2015,Isaule2016,Arellano2016}, 
to subsequently obtain fully off-shell $g$ matrices.

For purposes of this study it has been crucial to rely on
accurate means to obtain
scattering observables in the presence of nonlocal potentials, 
including the long range Coulomb interaction. 
This is achieved with the use of recently released packages
{\small SWANLOP}: 
{Scattering WAves off NonLocal Optical 
Potentials}~\cite{Arellano2021,Arellano2019}; and
{\small SIDES}:
{Schr\"odinger Integro-Differential Equation
Solver}~\cite{Blanchon2020}.
Both packages become suited for nucleon scattering off light and
heavy targets, at beam energies ranging from a few MeV up to 1~GeV.
No conditions are made to the local/nonlocal structure of the hadronic
part of the potential, as long as it is finite range.

\section{Calculations and findings}
\label{sec:findings}

We begin by exploring the systematics of the 
resulting scattering observables under varying cutoff momenta
$\Lambda$, covering a broad range of target masses $A$ and
beam energy $E_{Lab}$.
To this purpose, we calculate optical model potentials for proton elastic 
scattering off $^{4}$He, $^{16}$O, $^{40}$Ca, $^{90}$Zr, and $^{208}$Pb.
Eleven beam energies 
are considered: 30~MeV, and
from 100~MeV up to \num{1}~GeV in steps of 100~MeV.
In this case all $g$ matrices are based on AV18 \textit{NN}
bare potential.
Applications for 400~MeV and above include 
a non-Hermitian separable term added to AV18 reference potential
in order to reproduce exactly \emph{NN} scattering 
amplitudes above pion production threshold~\cite{Arellano2002}. 
The one-body target mixed density is represented in the
Slater approximation~\cite{Arellano1990}, for which we only need
radial point densities for protons and neutrons.
In this case we use densities described in Ref.~\cite{Aguayo2008}.
The momentum array for $\tilde U(k',k)$ is set as
$0\leq k\leq {K}$, with
$K=\max (8\textrm{ fm}^{-1},2k_0)$, where 
$k_0$ is the relative momentum in the 
\textit{NA} center-of-momentum (c.m.) reference frame.

\begin{figure} [ht]
  \begin{center}
  \includegraphics[width=\linewidth,angle=00]{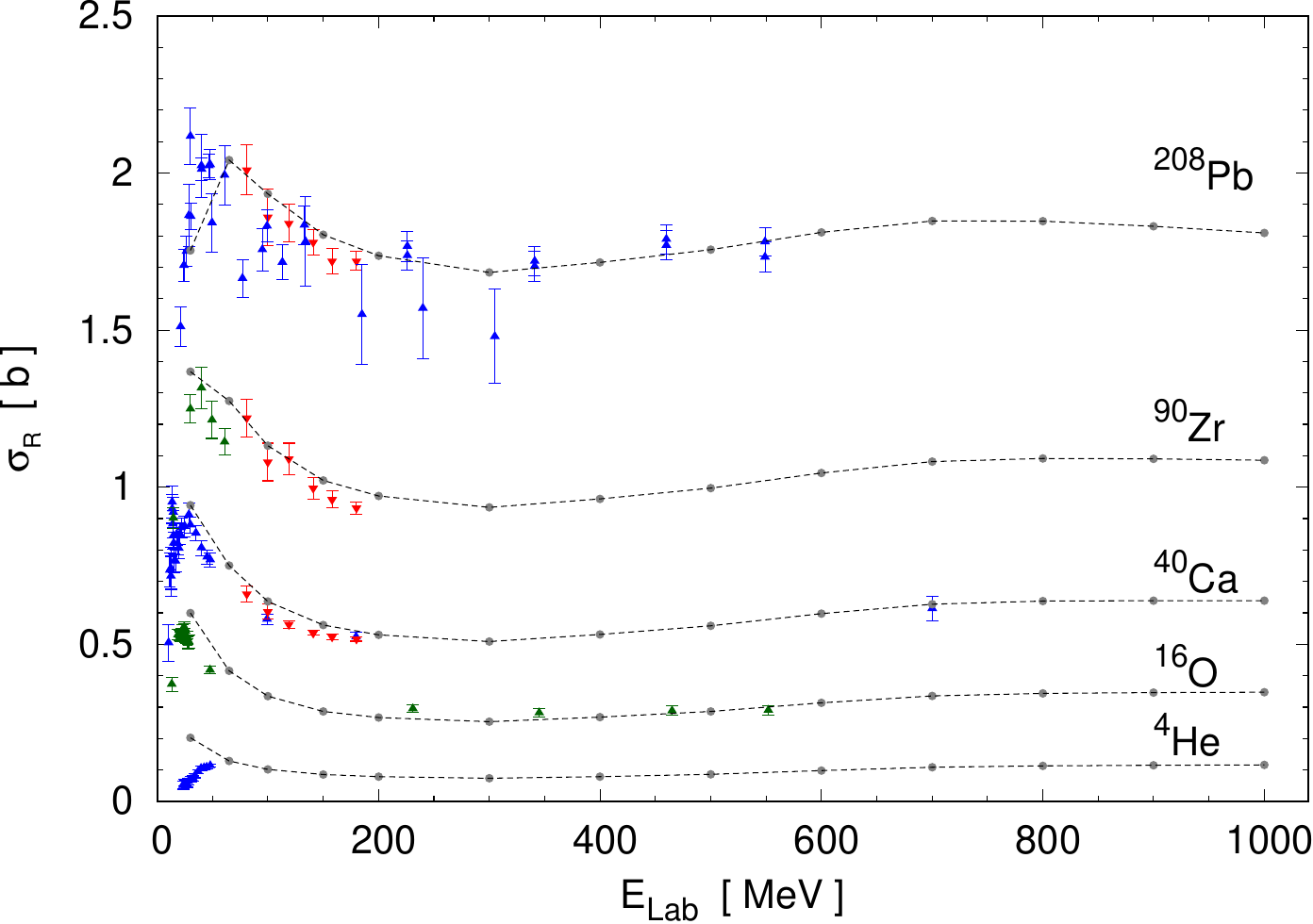}
  \caption{
  Reaction cross section for proton-nucleus elastic scattering scattering 
  as a function of the beam energy. Targets include
  $^{4}$He, $^{16}$O, $^{40}$Ca, $^{90}$Zr, and $^{208}$Pb.
  Downward and upward triangles denote data from 
  Refs.~\cite{Auce2005} and \cite{Carlson1996}, respectively.
          }
  \label{fig:sigmaR}
  \end{center}
\end{figure}

In Fig.~\ref{fig:sigmaR} we show the resulting
total reaction cross section as a function of the 
beam energy for proton elastic scattering off
$^{4}$He, $^{16}$O, $^{40}$Ca, $^{90}$Zr, and $^{208}$Pb.
Filled circles denote actual results from the optical model, 
with short-dashed curves included to guide the eye.
Downward red triangles denote data from Ref.~\cite{Auce2005}.
Blue and green upward triangles represent data
from Ref.~\cite{Carlson1996}, with datum for $^{208}$Pb$(p,p)$ at
860 MeV excluded as it corresponds to attenuation
cross section~\cite{Lapoux2017,Chen1955}.
We observe reasonable agreement between the calculated cross sections
and the data over a broad energy range ($\sim\!1$~GeV), 
validating the soundness of the optical model used in this study.
%==================================================================
%==================================================================

\subsection{Invariant sector}
\label{sec:sector}
We now investigate the role of high momentum components for
the description of the scattering process, specifically
its associated scattering observables.
Thus, we look for a minimum cutoff momentum $\Lambda$ which
guarantees accurate results for the total reaction cross section. 
In Ref.~\cite{Arellano2018} this study was limited to
$p+^{40}$Ca scattering,
obtaining that the minimum cutoff follows the rule
$\Lambda^2\! =\! \Lambda_0^2\!+\!k_E^2$, 
with $\Lambda_0\!=\!2.4$~fm$^{-1}$, 
and $k_E$ the momentum of the projectile in the laboratory reference frame.
We aim here to extend that result by considering the cases
$A=4$, 16, 40, 90, and 208. We proceed as follows.

\begin{figure}  [ht]  
  \includegraphics[width=0.95\linewidth,angle=0]{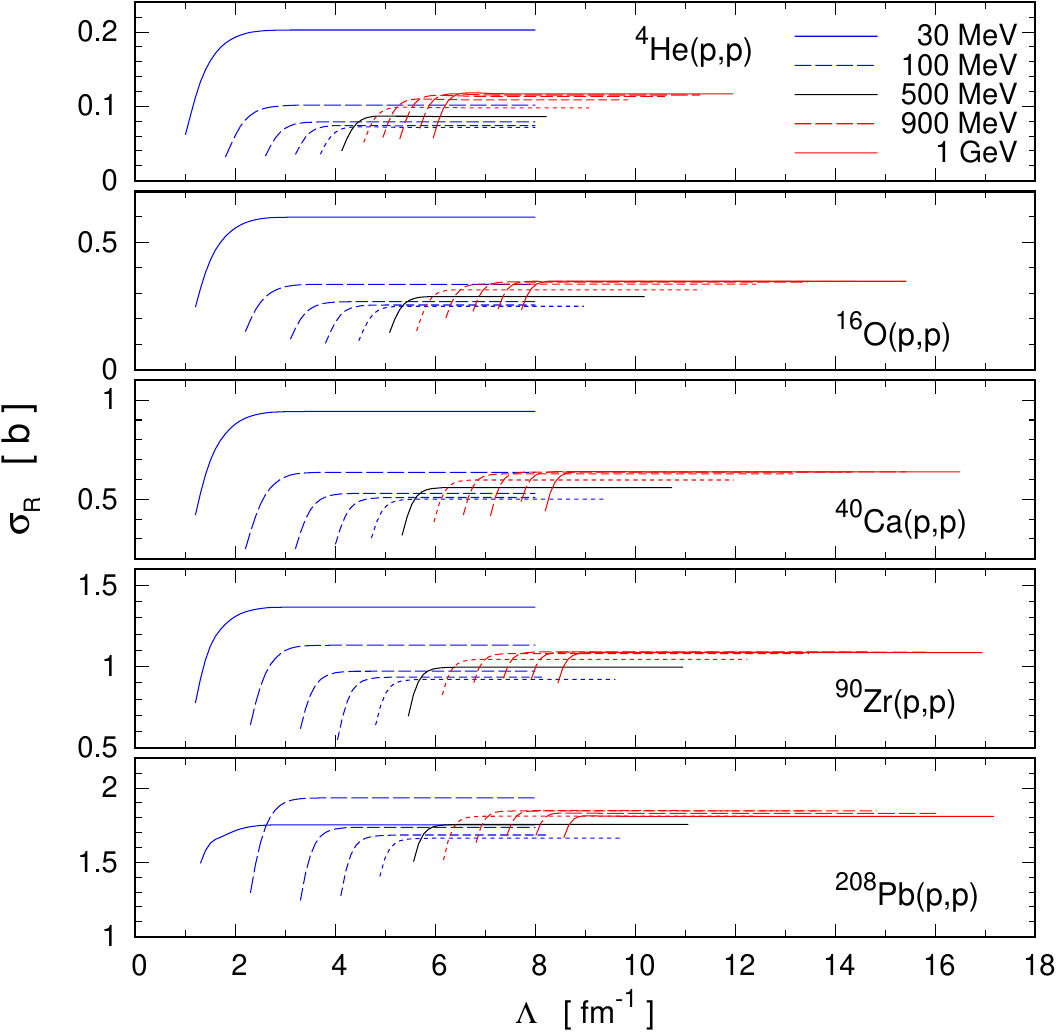}
  \caption{
Reaction cross section for proton-nucleus scattering as a function of the 
cutoff momentum $\Lambda$ applied to momentum-space optical potentials.
Targets considered are 
  $^{4}\textrm{He}$,
  $^{16}\textrm{O}$,
  $^{40}\textrm{Ca}$,
  $^{90}\textrm{Zr}$ and 
  $^{208}\textrm{Pb}$,
at energies between 30~MeV and 1~GeV.
See text for description of curve patterns.
          }
  \label{fig:x5}       % Give a unique label
\end{figure}

For a given target and energy we calculate $\sigma_R$ for a sequence
of cutoff momenta $\Lambda_i$, starting from $\Lambda_1=K$ and
ending whenever $\Lambda_i$ is at or below the relative momentum
in the c.m. reference frame.
The spacing between consecutive values of $\Lambda$ is
$\delta\Lambda=0.1$~fm$^{-1}$.
In this way the calculated reaction cross section, 
$\sigma_i=\sigma(\Lambda_i)$, 
will depend on the target mass number $A$, the beam energy $E$ and 
the cutoff momentum $\Lambda_i$.
In Fig.~\ref{fig:x5} we show the resulting reaction cross section
as a function of the cutoff momentum $\Lambda$.
Each curve corresponds to a specific energy.
Blue curves denote results for $E=30$, 100, 200, 300 and 400~MeV;
solid black curves denote results for $E=500$~MeV; and
red curves represent results for 
$E=600$, 700, 800, 900 and \num{1000}~MeV.
As observed, 
all cases exhibit a plateau above a given cutoff momentum.
%==================================================================
%==================================================================

\begin{figure} [ht]
  \includegraphics[width=0.9\linewidth,angle=0]{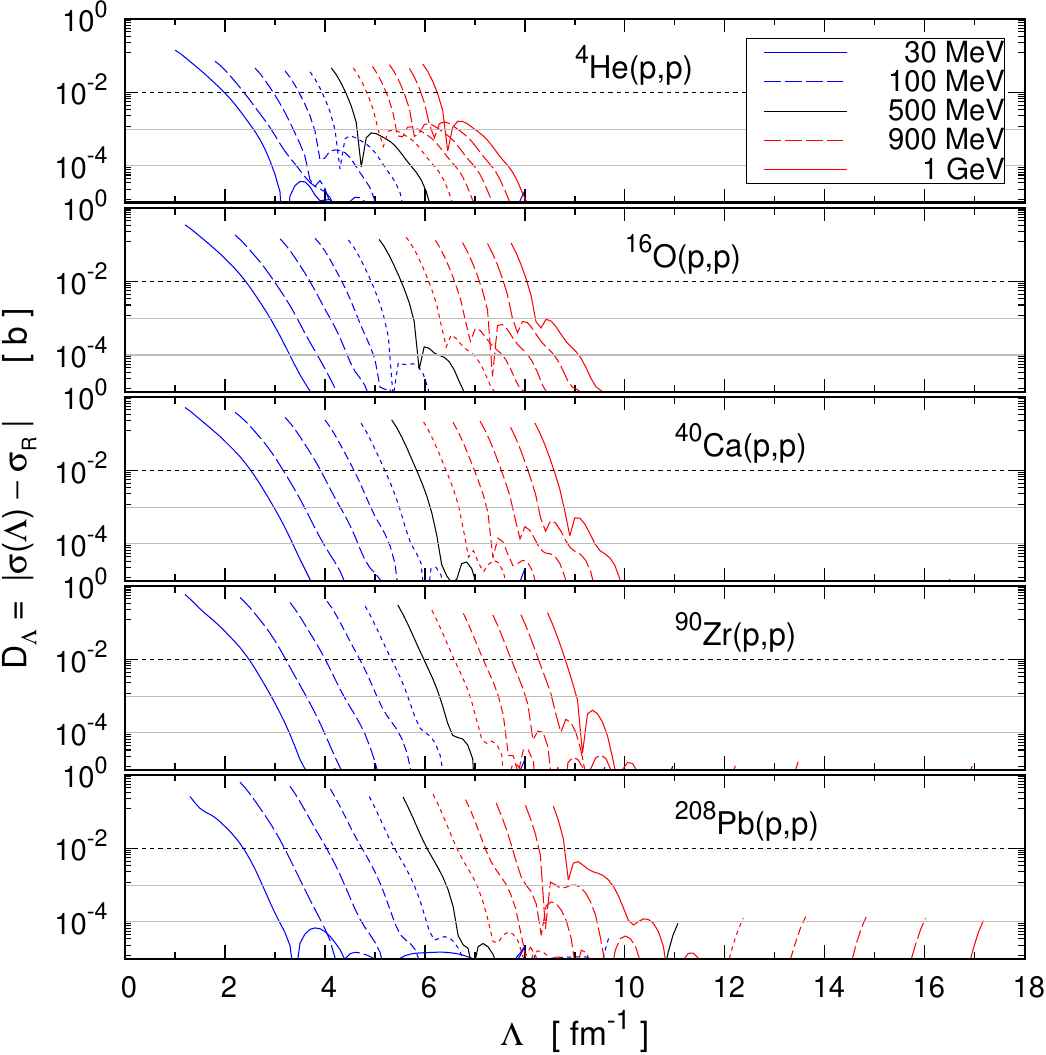}
  \caption{
  Departure from the plateau-value of the calculated reaction cross 
  section as a function of $\Lambda$.
  Curve patterns follow the same convention as in Fig.~\ref{fig:x5}.
          }
  \label{fig:gradients}       % Give a unique label
\end{figure}

In order to identify the threshold cutoff momentum $Q$ we scrutinize 
the cross section at the plateau. 
We first calculate the plateau-value cross section $\sigma_R$,
which we define as the average at the plateau considering $\sigma_i$ 
whose forward gradient
$|(\sigma_{i+1}-\sigma_{i})|/\delta\Lambda$, 
is smaller than $10^{-4}$~{b~fm}.
In Fig.~\ref{fig:gradients} we show logarithmic 
plots of the absolute difference 
$D_\Lambda\!=\!|\sigma(\Lambda) - \sigma_R|$,
as a function of the cutoff $\Lambda$ for the five targets considered.
Curve patterns and colors follow the same convention
as those in Fig.~\ref{fig:x5}.
We note that the differences $D_\Lambda$ decrease 
sharply with the cutoff momentum.
Based on the steep descent of $D_\Lambda$, we define the threshold 
cutoff momentum $Q$ as that where the absolute error with respect
to the plateau average crosses 10$^{-2}$~b. 
With this criterion we obtain a well defined estimate of the minimum 
$\Lambda$ at which the calculated cross sections does not
change within the specified accuracy. 
%==================================================================
%==================================================================

\begin{figure} [ht]
  \begin{center}
  \includegraphics[width=0.9\linewidth,angle=0]{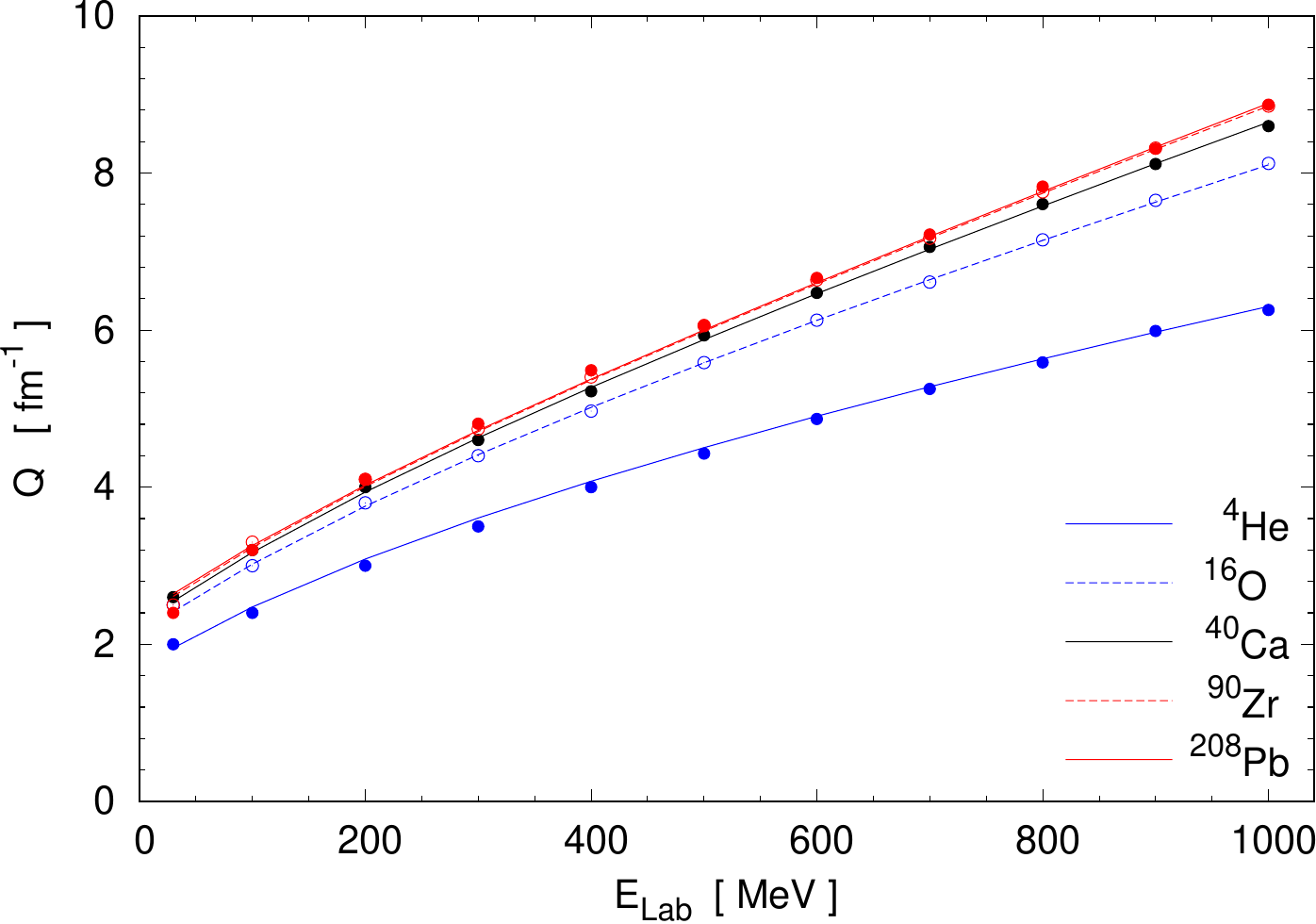}
  \caption{
  Threshold cutoff momentum $Q$ as a function of the beam energy for 
  proton scattering off selected targets.
  Solid curves represent the parametrization given by Eq.~\eqref{xoff}.
          }
  \label{fig:lambda}
  \end{center}
\end{figure}

In Fig.~\ref{fig:lambda} we plot with circles
the obtained threshold cutoff momentum $Q$ as a function
of the beam energy $E_{Lab}$ for the five targets under consideration.
We note that $Q$ increases with the beam energy
and the target mass number $A$. 
We have found a 
simple parametrization for the observed behavior, summarized by
\begin{equation}
  \label{xoff}
  Q = \sqrt{a^2 + b\;k^2}\;,
\end{equation}
with $k$ the relative momentum in the \textit{NA} c.m.
reference frame. 
Here $a$ and $b$ depend on the target mass number $A$ 
as follows:
\begin{subequations}
 \begin{align}
   \label{aa}
  a =&\frac{3}{5}\left(4 -\frac{3}{A^{2/3}}\right)~\textrm{fm}^{-1}\;;
    \\
    \label{bb}
   b =& \frac{1.05}{1+\num{1.7e-4}A} \;.
    \end{align}
\end{subequations}
Results from this parametrization are shown with continuous curves
in Fig.~\ref{fig:lambda}, where we observe a close correspondence
with the calculated $Q$ shown with circles. 
%==================================================================
%==================================================================

\begin{figure} [ht]
  \includegraphics[width=0.95\linewidth,angle=0]{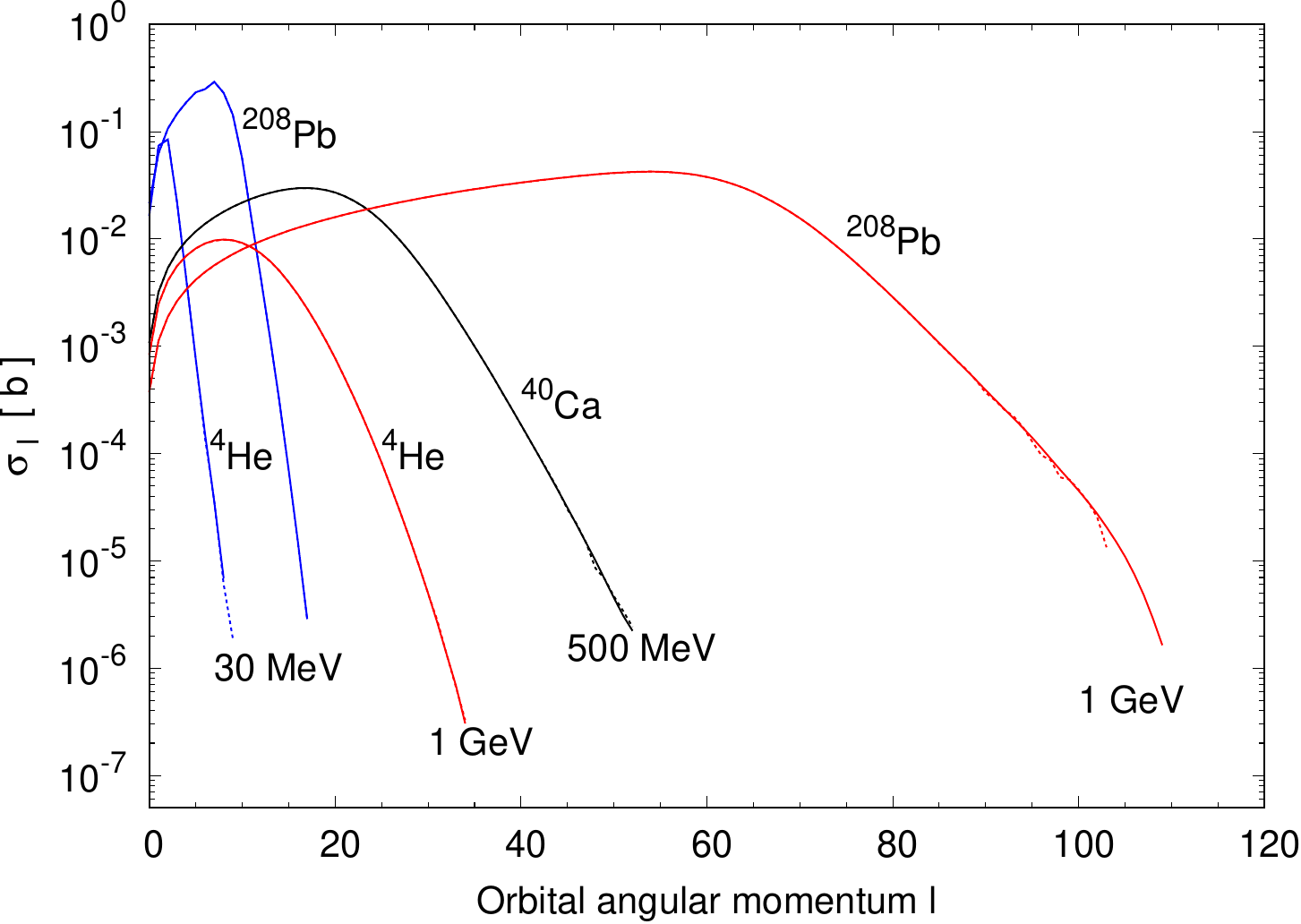}
  \caption{
    Partial absorption $\sigma_l$ for proton-nucleus scattering
    as functions of partial waves.
Blue, black and red curves denote results at
30~MeV, 500~MeV and 1~GeV, respectively.
  Solid curves use $\Lambda\!=\!Q\!+\!1$~fm$^{-1}$, from Eq.~\eqref{xoff},
  while dotted curves use $\Lambda=K$.
          }
  \label{fig:orbital}       % Give a unique label
\end{figure}

It is worth stressing that the calculated $Q$ delimits
a boundary beyond which there is no meaningful physical content in the
potential. 
This threshold is not set \textit{a priori} but stems from 
a convergence criterion on the calculated cross sections.
Any cutoff below this threshold alters the calculated observables.
Conversely, whenever the cutoff is above the boundary, cross sections
become invariant.
This feature is illustrated in Fig.~\ref{fig:orbital}, 
where we plot the partial cross section
\begin{equation}
  \label{sigmal}
  \sigma_l = \frac{\pi}{k^2} 
  \left [
%   (l+1) (1 - |S_{l-\!\textstyle{\frac12},l}|^2 ) +
    (l+1) (1 - |S_{l-1/2,\,l}|^2 ) +
     l    (1 - |S_{l+1/2,\,l}|^2 ) 
     \right ]\;,
\end{equation}
as a function of the orbital angular momentum $l$.
Here $S_{jl}=\exp(2i\delta_{jl})$, 
with $\delta_{jl}$ the phase-shift for total and orbital angular
momentum $j$ and $l$, respectively.
Blue, black and red curves denote results at
30~MeV, 500~MeV and 1~GeV, respectively.
Results for $^{4}$He, $^{40}$Ca and $^{208}$Pb are included.
Solid curves correspond to results using 
$\Lambda\!=\!Q\!+\!1$~fm$^{-1}$, to move away from the transient.
Dotted curves have been taken using $\Lambda=K$, the maximum momentum
at which the potential has been evaluated.
We observe near complete overlap between solid and dashed curves, 
with the
exception of high $l$ in the case of $^{40}$Ca at 500~MeV, and
$^{208}$Pb at 1~GeV. 
We have found that those fluctuations are due to the exceedingly
high $K$ in both cases. 
The fluctuations disappear if we limit $K$ to 12~fm$^{-1}$. 
%==================================================================
%==================================================================

\subsection{Momentum- and coordinate-space structure}
\label{sec:structure}

\begin{figure} [ht]
  \begin{center}
  \includegraphics[width=0.9\linewidth,angle=0]{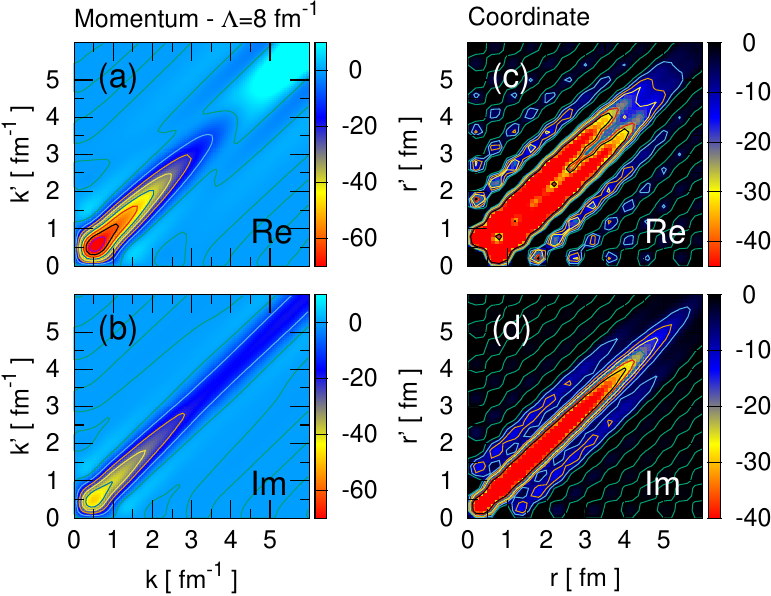}
  \caption{
   $s$-wave optical potential based on AV18 for $p+^{40}$Ca 
   scattering at 65~MeV.
   LHS (RHS) panels show potential in momentum (coordinate) representation.
   Upper (lower) frame show real (imaginary) part.
   Case for $\Lambda=8$~fm$^{-1}$.
          }
  \label{fig:u700}
  \end{center}
\end{figure}

Momentum-space potentials in their general form
have the advantage of retaining naturally
intrinsic nonlocalities. 
However, there are no studies relating their coordinate-space
structure with well established models in coordinate space.
Let us consider $p+^{40}$Ca elastic scattering with proton beam
energy of 65~MeV.
In this case we consider a momentum-space optical potential 
based on AV18 bare \textit{NN} interaction.
On the left-hand side (LHS) of Fig.~\ref{fig:u700} we show 
contour plots for the real (a) and imaginary (b)
$s$-wave potential $k'\tilde U(k',k)k$.
The corresponding coordinate-space real and imaginary parts
of $r'U(r',r)r$ are shown in the right-hand side (RHS) panels
(c) and (d).
The momentum-space potential is calculated with
$K\!=\!8$~fm$^{-1}$.
For clarity in the plots, the imaginary part of the potential
has been multiplied by a factor of two ($\times 2$).

We note that the momentum-space potential exhibits a smooth
behavior with its dominant
real and imaginary contributions along a diagonal band, with 
widths of about $1.5$ and $1$~fm$^{-1}$, respectively.
Their corresponding coordinate-space representation 
gets notoriously more structured, 
as evidenced with the sharp oscillatory patterns
in panels (c) and (d).
The dominant contributions in coordinate-space take place
near the diagonal.
%==================================================================
%==================================================================

\begin{figure} [ht]
  \begin{center}
  \includegraphics[width=0.9\linewidth,angle=0]{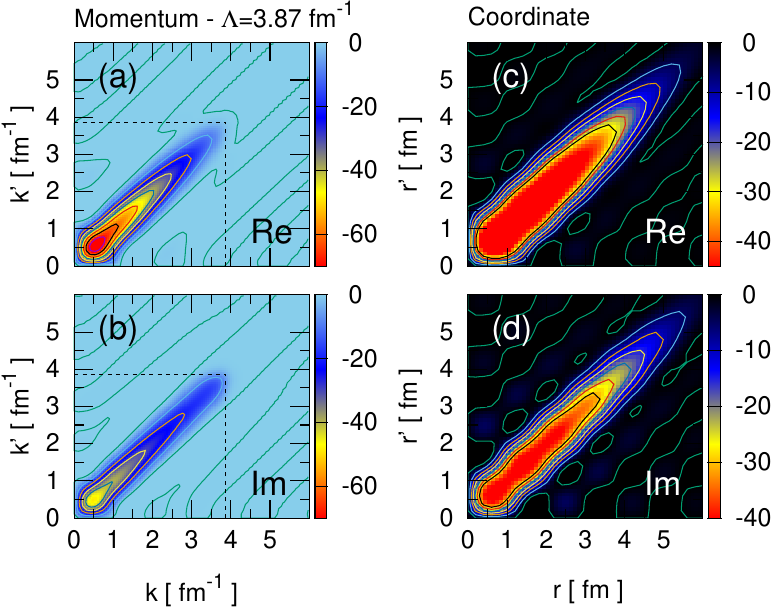}
  \caption{
    The same as in Fig.~\ref{fig:u700}, 
    but with $\Lambda=3.87$~fm$^{-1}$.
          }
  \label{fig:u387}       % Give a unique label
  \end{center}
\end{figure}

We now apply momentum cutoff to the above potential 
at a minimum $\Lambda$ which assures to account for its associated
scattering observables. 
To this purpose we take $Q$ from Eq.~\eqref{xoff}, 
adding $1$~fm$^{-1}$ in order to move away from the transient.
The resulting potentials are shown in Fig.~\ref{fig:u387},
where we use the same scales and conventions as in Fig.~\ref{fig:u700}.
In this case panels (a) and (b) for the momentum-space potential
evidence the suppression of momentum components above $\Lambda$.
As a result,
its corresponding coordinate-space representation becomes less structured,
with a clear and smooth distribution away from the diagonal.
This extension off the diagonal in coordinate space evidences 
nonlocality of the interaction.
Beyond the drastic differences between coordinate-space potentials
shown in Figs.~\ref{fig:u700} and \ref{fig:u387}, we verify that
all \textit{NA}
scattering observables and wavefunctions are identical within numerical
accuracy.
%==================================================================
%==================================================================

\subsection{Assessment of nonlocality}
\label{sec:pbuck}

Thus far we have only considered momentum-space potentials and their 
resulting coordinate-space representation after suppression of ultraviolet
Fourier components. Cutoffs are applied in momentum space.
We now investigate coordinate-space models.
The idea in this case is to transform them into momentum space applying
a Fourier transform (FT), 
followed by a momentum cutoff at a given $\Lambda$,
and then transform back to coordinate space (FT$^{-1}$).
This procedure is illustrated in Fig.~\ref{fig:r2k2r}.
\begin{figure} [ht]
\begin{center}
\begin{tikzpicture} 
   \node[] (x0) at (0,0) {$U(r',r)$};
    \node[] (k0) at (3,0) {$\tilde U(k',k)$};
     \node[] (k1) at (3,-1.2) {$\tilde U_\Lambda(k',k)$};
      \node[] (x1) at (0,-1.2) {$U_\Lambda(r',r)$};
       \draw[->] (x0) to node[above] {FT} (k0);
        \draw[->] (k0) to node[right] {$\Lambda$} (k1);
          \draw[->] (k1) to node[above] {FT$^{-1}$} (x1);
 \end{tikzpicture}
  \caption{ \label{fig:r2k2r}
  Momentum cutoff to a potential in coordinate space.
  }
\end{center}
\end{figure}
For the Fourier transform back to momentum space we use
Eq.~\eqref{kk2rr} and obtain
\begin{equation}
  \label{rr2kk}
  \tilde U_{l}(k',k) = \frac{2}{\pi}
  \int_0^\infty r'^2dr' 
  \int_0^\infty r^2dr\,
  j_l(k'r') U_{l}(r',r) j_l(kr)\;.
\end{equation}
Note that this expression enables us to include any kind of finite
range potential, even local ones. 
For the latter we use $rU_l(r',r)r\!=\!V(r)\delta(r-r')$,
with $\delta(r-r')$ the one-dimensional Dirac delta function and
$V(r)$ the usual local potential.
The suppression of the high momentum components of the local
potential results in a nonlocal one.

With the above considerations we analyze Perey-Buck nonlocal
potentials, using Tian-Pang-Ma (TPM) 
parametrization~\cite{Tian2015}.
We also include in this analysis 
Koning-Delaroche (KD) phenomenological local optical 
model~\cite{Koning2003}.
In this case we focus on $p+^{40}$Ca elastic scattering 
at 30.3~MeV in the laboratory reference frame.
The two phenomenological 
potentials will be compared with microscopic momentum-space 
potentials based on N3LO and AV18 bare \textit{NN} interactions.
This energy has been chosen because all four optical models 
become applicable.
In all cases the calculated scattering observables are obtained
with momentum cutoff $\Lambda=3.87$~fm$^{-1}$, 
obtained from Eq.~\eqref{xoff} with an increment of 1~fm$^{-1}$.

\begin{figure} [ht]
\begin{center}
  \includegraphics[width=0.9\linewidth,angle=0]{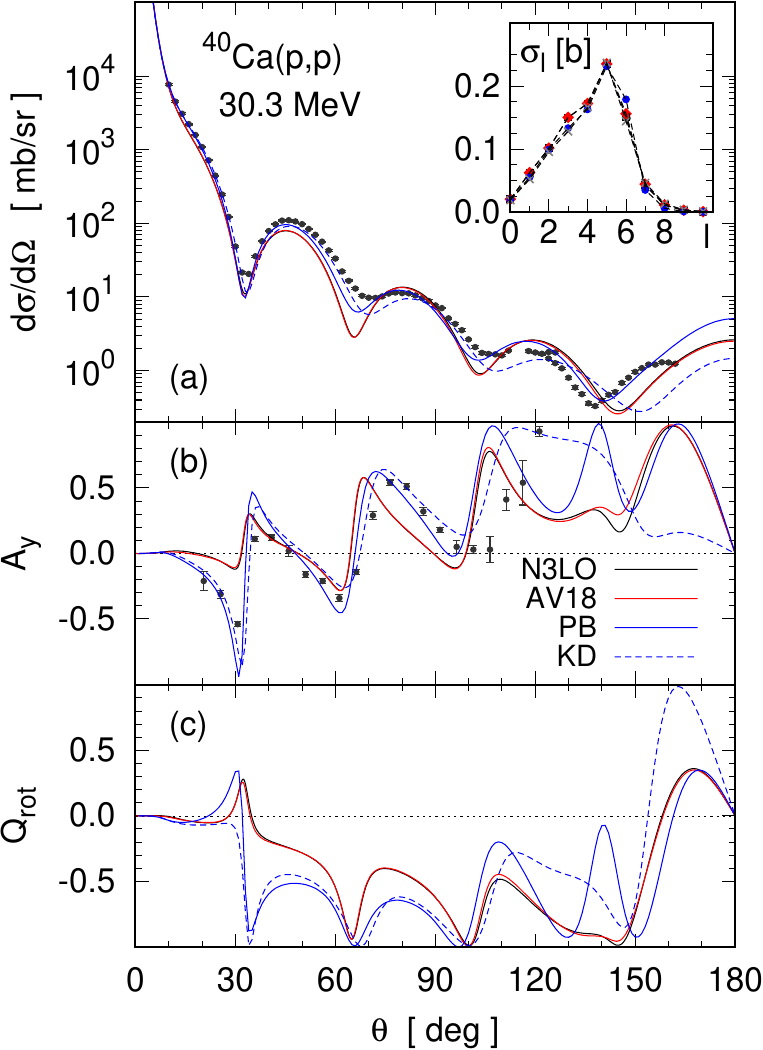}
  \caption{
Differential cross section 
$d\sigma/d\Omega$ (a),
analyzing power $A_y$ (b) and
  spin rotation function $Q_{rot}$ (c) as functions of the scattering
angle $\theta$ in the c.m. reference frame.
  Data from Ref.~\cite{Hnizdo1971}. 
See text for description of curve patterns.
Inset shows $\sigma_l$ as a function of the orbital
angular momentum $l$.
          }
  \label{fig:xaq}       % Give a unique label
\end{center}
\end{figure}

The ability of the four models to describe the data is shown 
in Fig.~\ref{fig:xaq},
where we plot the calculated differential cross section 
$d\sigma/d\Omega$ (a),
analyzing power $A_y$ (b) and
spin rotation function $Q_{rot}$ (c) 
as functions of the scattering
angle $\theta$ in the c.m. reference frame.
The data are from Ref.~\cite{Hnizdo1971}. 
The inset in (a) shows $\sigma_l$ as a function of the orbital
angular momentum $l$.
Results based on N3LO and AV18 potentials are denoted with
black and red curves, respectively.
Results for TPM parametrization and KD local potential
are shown with blue solid and dashed curves, respectively.
As observed, all models provide an overall reasonable description
of the data, with TPM and KD in closer agreement with measurements.
From this result we can state that all approaches contain the essential
elements for the description of the scattering process.
From the inset we also note that stronger absorption takes place for
$h$-waves ($l\!=\!5$), channel we shall pay attention to.
%==================================================================
%==================================================================

\begin{figure*} [ht]
  \begin{center}
  \includegraphics[width=0.95\linewidth,angle=0]{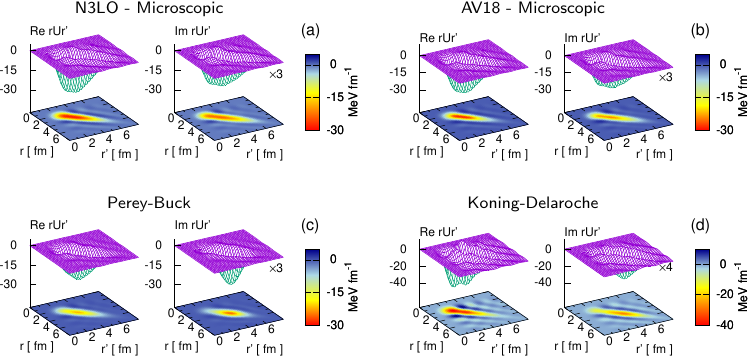}
  \caption{
  \label{fig:u2x2}       % Give a unique label
    Surface plots in coordinate space for $h$-wave 
    ($j\!=\!5\!+\!\nicefrac{1}{2}$) real (Re) and imaginary (Im) 
    potentials.
    Plots (a) are for the microscopic N3LO-based potential;
    (b) for microscopic AV18-based potential;
    (c) for Perey-Buck nonlocal potential with TPM parametrization; and
    (d) for Koning-Delaroche local potential.
    All potentials have momentum cutoff at $\Lambda=3.87$~fm$^{-1}$.
          }
  \end{center}
\end{figure*}

In Fig.~\ref{fig:u2x2} we show surface plots of $h$-wave
($j\!=\!l+\nicefrac{1}{2}$) potentials
in the $rr'$ plane. All potentials are subject to ultraviolet
cutoff $\Lambda=3.87$~fm$^{-1}$. 
Plots (a) represent results based on N3LO,
(b) for AV18,
(c) for Perey-Buck nonlocal model with TMP parameters,
and
(d) for Koning-Delaroche (KD) local potential.
The imaginary components have been amplified by three ($\times 3$)
in all cases except KD, where the amplification is four times
($\times 4$). 
We observe that all potentials exhibit similar shapes in coordinate
space, despite their different nature.
Indeed, the N3LO-based optical model is constructed from chiral
interactions with high momentum components already suppressed
at the \textit{NN} level. 
With this feature high Fourier components of the
$g$ matrix get suppressed, resulting in an
\textit{NA} potential more confined in momentum space.
Such is not the case of AV18, where high Fourier components are
present, extending the optical potential over the whole momentum domain.
In the case of the PB model, the definition of the potential in coordinate
includes Fourier components over the whole spectrum, 
which after ultraviolet cutoff get suppressed.
The same holds for KD local potential. Once transformed into
momentum space and suppressed its high Fourier components,
returns to coordinate space as nonlocal. 
What is remarkable from Fig.~\ref{fig:u2x2} is the close 
resemblance of all four potentials, despite their different origins.
%==================================================================
%==================================================================

\subsection{Transversal concavity}
\label{sec:concavity}
\begin{figure} [ht]
  \begin{center}
  \includegraphics[width=0.40\linewidth,angle=0]{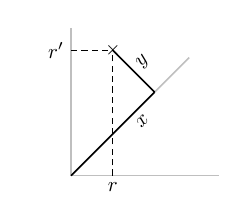}
  \caption{
  \label{fig:axes}       % Give a unique label
    Diagonal and transversal coordinates in $rr'$ plane.
          }
  \end{center}
\end{figure}

We now pay attention to the nonlocal structure of the
resulting potentials shown in Fig.~\ref{fig:u2x2}.
In particular, we focus on the transversal concavity of the potential
along the diagonal.
If the potentials were local, then
$rU(r,r')r'$ would vanish away from the diagonal
$r\!=\!r'$, becoming very strong along the diagonal.
To study these features let us introduce the alternative 
coordinate set $xy$ given by
\begin{equation}
  x  = {\frac{1}{\sqrt{2}}}( r'+ r ) \;;\qquad
  y  = {\frac{1}{\sqrt{2}}}( r'- r ) \;.
\end{equation}
As illustrated in Fig.~\ref{fig:axes}, 
this represents a forty-five degree counter-clockwise rotation
of the $rr'$ axes, with $y$ representing the departure from
the diagonal defined by $r\!=\!r'$. 
%==================================================================
%==================================================================
With these coordinates we denote
%\begin{equation}
%  \label{uxy}
${\cal U}(x,y) \equiv r'U_l(r',r)r$.
%\end{equation}
To examine the potential in vicinities of the diagonal we perform
a series expansion up to second order in the $y$ coordinate 
\begin{equation}
 \label{taylor}
 {\cal U}(x,y) = {\cal U}(x,0)+\frac12{\cal U}''(x,0)\,y^2 +
  {\cal O}(y^4)\;,
\end{equation}
with ${\cal U}''(x,0)\!\equiv\!\partial^2U(x,y)/\partial y^2|_{y=0}$, 
the concavity of the potential on the diagonal.

To guide an interpretation of the concavity of the 
potential in the $rr'$ plane, let us examine
Perey-Buck nonlocal construction.
In this model the central term takes the separable structure
\begin{equation}
    \label{pbuck}
      U({\bm r}',{\bm r})=V({\bm R})\,H({\bm s})\;,
\end{equation}
where
\begin{equation}
      {\bm R} = \textstyle{\frac12}({\bm r} + {\bm r}') \;; \qquad
          {\bm s} = {\bm r}' - {\bm r} \;.
\end{equation}
Form factor $V$ is complex of Woods-Saxon type, including a 
surface term. 
The $H$ form factor allows for nonlocality, given by a 
normalized Gaussian of width $\beta$ expressed as
\begin{equation}
    \label{gauss}
      H({\bm s}) = \frac{1}{{\pi}^{3/2} \beta^3} e^{-s^2/\beta^2}\;.
\end{equation}
Parameter $\beta$ is commonly used to gauge degree
of nonlocality in some studies.

To obtain the $l$-th multipole of the potential we evaluate
\begin{equation}
  \label{pbl}
    U_{l}(r',r) = 2\pi\int_{-1}^{1} P_{l}(u)\; V(R)\,H(s) \,du\;,
\end{equation}
where $u\!=\!\hat r\cdot\hat r'$. 
Since $H(s)$ is sharply peaked for $s\!\approx\!0$,
then leading contributions from $V$ take place 
at $R\!\approx\! x/\sqrt{2}$. 
If we denote ${\cal U}_{PB}(x,y)=r'U_{l}(r',r)r$,
some direct simplifications yield
\begin{equation}
  \label{ulrr}
  {\cal U}(x,y) \approx
  \frac{2}{{\pi}^{1/2}\beta^3}
    V\left(\frac{x}{\sqrt{2}}\right )\,
    e^{-(x^2+y^2)/\beta^2} 
   \, w_l \left( \textstyle{\frac{x^2-y^2}{\beta^2}} \right )\;,
\end{equation}
where 
\begin{equation}
    \label{wl}
      w_l(b)={b}\int_{-1}^{1} P_l(u)e^{bu}\,du\;.
\end{equation}
In Appendix~\ref{awwll} we provide closed expressions for
$w_l(b)$ in the cases $l\leq 5$, being expressed as
\begin{equation}
  \label{wwll}
    w_l(b) = e^{b} {y}_l \left (\textstyle{\frac{-1\,}{b}} \right ) 
           -e^{-b} {y}_l \left (\textstyle{\frac{\,1\,}{\,b\,}} \right )\;,
\end{equation}
with ${y}_l(b)$ Bessel polynomials of order $l$.
Upon substitution into Eq.~\eqref{ulrr}, 
after Taylor expansion in the transversal coordinate $y$, we obtain
\begin{equation}
    \label{taylorPB}
  {\cal U}_{PB}(x,y) \approx
        \frac{2 V(x/\sqrt{2})}{{\pi}^{1/2}\beta^3}
        \left [ 
            1 - \frac{2y^2}{\beta^2(1-e^{-2x^2/\beta^2})} + 
            {\cal O}(y^4)
        \right ]\;.
\end{equation}
The term accompanying $y^2$ represents the acuteness of the potential
along the diagonal, providing a quantitative measure of nonlocality.
A comparison of this approximate result with that in Eq.~\eqref{taylor} 
leads us to introduce $\kappa$, a measure of nonlocality and defined by
\begin{equation}
  \label{qwidth}
  {\kappa} = - 4\frac{{\cal U}(x,0)}{{\cal U}''(x,0)}\;.
\end{equation}
In the case of approximation in Eq.~\eqref{ulrr} for PB we obtain
\begin{equation}
  \label{kappaPB}
   \kappa_{PB} \approx (1-e^{-r^2/\beta^2}) \,\beta^2\;,
\end{equation}
which for $r\gg\beta$ along the diagonal converges to $\beta^2$,
the square of PB nonlocality parameter.
In general, $\kappa$ is channel dependent.

\begin{figure} [ht]
  \begin{center}
  \includegraphics[width=0.9\linewidth]{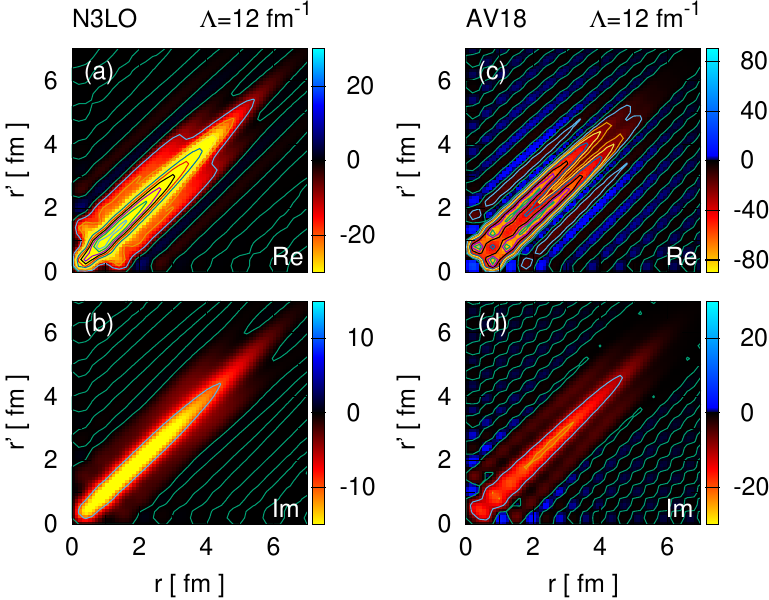}
  \caption{
    Contour plots $s$-wave optical potential in coordinate space
    obtained from momentum-space calculations using N3LO (LHS panels)
    and AV18 (RHS panels) \textit{NN} models.
    Momentum cutoff at $\Lambda\!=\!12$~fm$^{-1}$.
    Color bar in MeV~fm$^{-1}$ units.
          }
  \label{fig:rspace-s}       % Give a unique label
  \end{center}
\end{figure}

In Fig.~\ref{fig:rspace-s} we show surface plots of $s$-wave potentials
$rU(r,r')r'$ in the $rr'$ plane. 
We include microscopic potentials based of leading-order bare potential
N3LO (LHS panels) and AV18 (RHS panels).
The real parts of the potentials are shown in frames (a) and (c),
respectively.
Their corresponding imaginary parts are shown in panels (b) and (d).
Both potentials are constructed in momentum space, with
$\Lambda=12$~fm$^{-1}$.
As in the case of $h$ waves at 65~MeV, the coordinate-space potential
is much structured and stronger in the case of AV18 than for N3LO.
Observe the $[-80\!:\!80]$~MeV~fm$^{-1}$ scale in panel (c) for AV18, 
in contrast with $[-20\!:\!20]$~MeV~fm$^{-1}$ scale in panel (a) for N3LO.
%==================================================================
%==================================================================

From the above results we can now evaluate $\kappa$. 
In this case we treat separately the real and imaginary parts
of the potential, leading to their respective $\kappa_R$ and
$\kappa_I$.
In Fig.~\ref{fig:beta2oo} we plot results for 
$\kappa_R$ (solid curves) and $\kappa_I$ (dashed curves) 
as functions of $r$.
Panels (a)  and (b) show results for $s$ and $h$ waves, respectively.
Black and red curves represent results for N3LO- and AV18-based 
microscopic potentials, respectively.
Blue curves correspond to the PB-TPM nonlocal model.
Dotted curves correspond to $\kappa_{PB}$ as in Eq.~\eqref{kappaPB}.
The solid horizontal line represents $\beta^2$, with 
$\beta=0.88$~fm$^{-1}$, from TPM parametrization.
We can state the following observations:
%==================================================================
\begin{figure} [ht]
  \begin{center}
  \includegraphics[width=0.8\linewidth,angle=0]{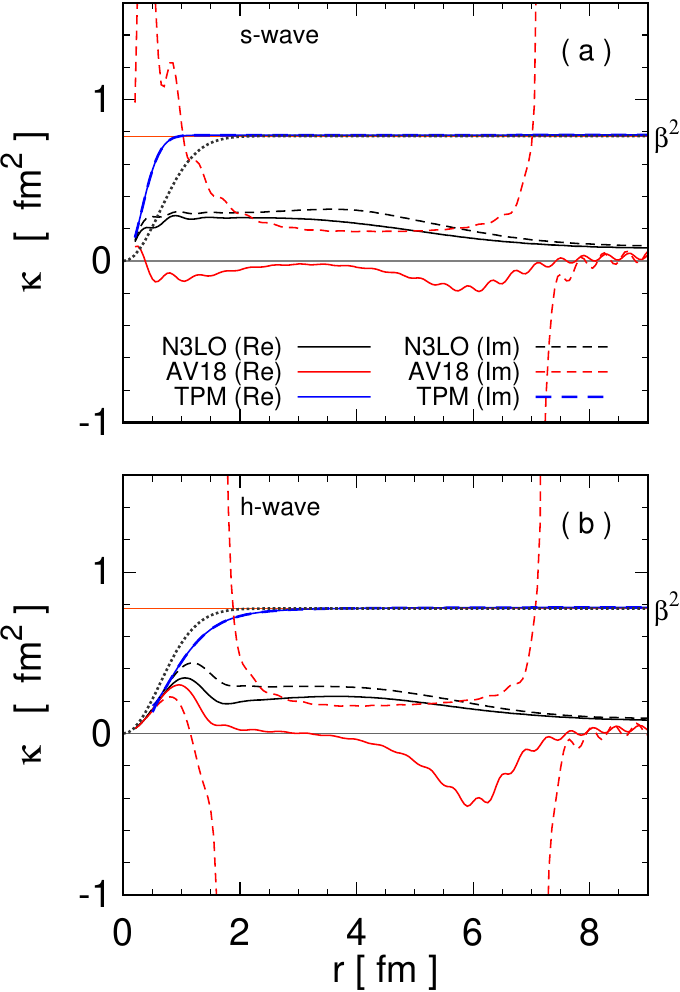}
  \caption{
    Diagonal $\kappa_R$ and $\kappa_I$ as functions of $r$. 
    Panel (a) shows results for $s$ wave, whereas panel (b) for
    $h$ wave.
    Black and red curves correspond to N3LO- and AV18-based microscopic
    potentials, respectively.
    Blue curves correspond to the PB-TPM nonlocal model.
    Solid and dashed curves represent $\kappa_R$ and $\kappa_I$, 
    respectively.
    Dotted curves correspond to approximations in Eq.~\eqref{kappaPB}.
          }
  \label{fig:beta2oo}       % Give a unique label
  \end{center}
\end{figure}
%==================================================================
\begin{enumerate}[label=\alph*)]
  \item
Black solid and dashed curves (N3LO-based) for $s$ waves 
are smooth and positive, showing similar behavior for $\kappa_R$ and 
$\kappa_I$. The same feature holds for the $h$ wave. 
The fact that these values for $\kappa$ are a fraction of $\beta^2$ 
indicates that the potential is sharper than the PB model
along the diagonal.
\item
Red solid curves (AV18-based) appear more irregular than all the
other cases. There is a change of sign which,
after a close inspection of panel (c) in Fig.~\ref{fig:rspace-s}, 
can be attributed to change of sign of the potential.
In the case of the imaginary part (red dashed curves) we note
singularities in $\kappa$, feature due to vanishing 
${\cal U}''(x,0)$ (real or imaginary components) along the diagonal.
\item
Solid and dashed blue curves (PB model) overlap completely, 
corresponding to $\kappa_R$ and $\kappa_I$, respectively.
Additionally, they show a smooth and uniform behavior, 
reaching a near constant value for $r$ above
    $1$~fm ($s$ wave) and $2.5$~fm ($h$ wave).
\end{enumerate}
A main conclusion from the preceding analysis is that all three 
potentials appear very different from one another 
when represented in coordinate space.
This is particularly the case of N3LO- vs AV18-based potentials,
where $\kappa_R$ and $\kappa_I$ behave quite differently.
This scenario changes radically with the suppression of ultraviolet 
components of \textit{NA} potentials, as we shall see next.

\begin{figure} [ht]
  \begin{center}
  \includegraphics[width=0.9\linewidth,angle=0]{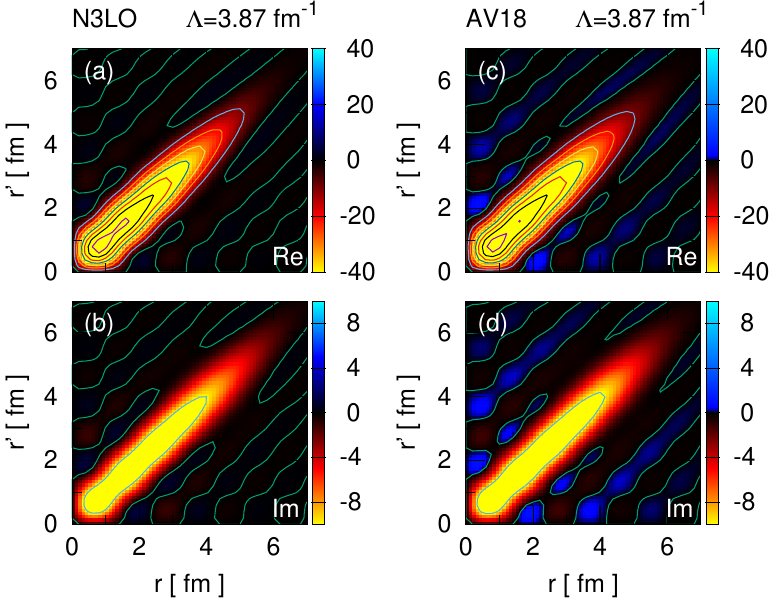}
  \caption{
    The same as in Fig.~\ref{fig:rspace-s}, but with
    $\Lambda\!=\!3.87$~fm$^{-1}$.
          }
  \label{fig:rspace-sx}       % Give a unique label
  \end{center}
\end{figure}

Considering the same potentials as above we proceed to suppress 
momentum components beyond $\Lambda\!=\!Q+1$~fm$^{-1}$. 
This is done directly to the N3LO- and AV18-based microscopic 
optical potentials. 
The resulting $s$-wave coordinate-space potentials are shown in 
Fig.~\ref{fig:rspace-sx}, whose description is the same as
for Fig.~\ref{fig:rspace-s}.
The only difference in this case is that the color bar range
in frames (a) and (c) are now the same.
Observe that the suppression of high momentum components 
in both cases results in potentials very similar to one another.
%==================================================================
%==================================================================

\begin{figure} [ht]
  \begin{center}
  \includegraphics[width=0.8\linewidth,angle=0]{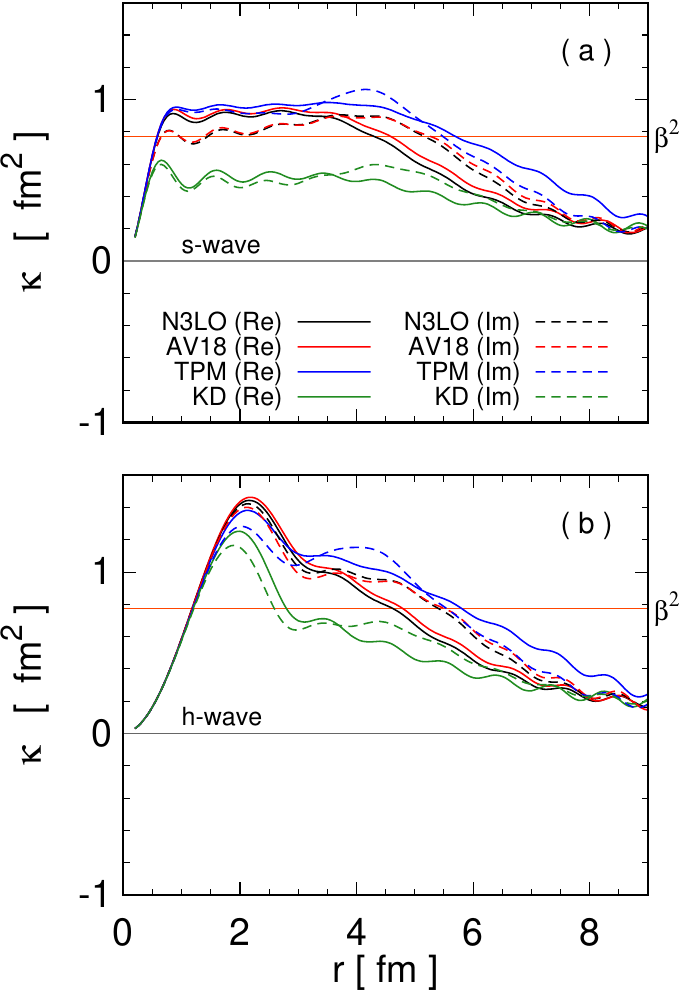}
  \caption{
    The same as in Fig.~\ref{fig:beta2oo}, but with
    $\Lambda\!=\!3.87$~fm$^{-1}$.
    Green curves denote $\kappa_R$ (solid) and $\kappa_I$ (dashed) 
    for KD potential.
          }
  \label{fig:beta2xx}       % Give a unique label
  \end{center}
\end{figure}

We can now examine the transversal concavity of the resulting potentials. 
In this analysis we also include Perey-Buck potential as well as 
Koning-Delaroche local model, both with their momentum components
above $\Lambda$ suppressed.
In Fig.~\ref{fig:beta2xx} we plot $\kappa$ as a function of $r$
for N3LO- and AV18-based microscopic optical potentials 
(black and red curves, respectively), as well as 
those based on
Perey-Buck nonlocal model (blue curves). 
Results for Koning-Delaroche potential are shown with green curves.
Solid and dashed curves correspond to $\kappa_R$ and 
$\kappa_I$, respectively.
Frames (a) and (b) show results for $s$ and $h$ wave, respectively.
%==================================================================
%==================================================================
In contrast to $\kappa$ in the cases with no suppression of
high momentum components in the potential, results shown in
Fig.~\ref{fig:beta2xx} show a smoother and less irregular behavior.
Indeed, we note that N3LO- and AV18-based microscopic potentials
lead to similar $\kappa_R$ and $\kappa_I$, 
in both $s$ and $h$ waves.
Additionally, these two models yield comparable $\kappa$ in the bulk
of the nucleus ($r\lesssim 3.5$~fm$^{-1}$). 
At the surface, the PB model behaves more nonlocal than microscopic ones.
In the case of KD potential, 
the resulting nonlocality as given by $\kappa$ is smaller than
for all the other cases, feature which appear more pronounced
in the case of $s$ waves.

\subsection{Discussion}
\label{sec:discussion}

We have identified a threshold momentum $Q$
that separates the low-momentum scale of the optical model potential
from the high-momentum components. 
We have shown that those high momentum components
become irrelevant for the evaluation of associated elastic
scattering observables. 
We stress that the threshold momentum $Q$ is not
set \textit{a priori} but inferred in the context of
realistic constructions of optical model potentials.
The criterion is that of being the smallest 
momentum window that enables one to reproduce the scattering 
observables within a given numerical accuracy.
On this regard,
the philosophy of the approach differs from that of renormalization group 
techniques for the constructions of $v$ low-$k$ \textit{NN} interactions,
where a momentum cutoff is set beforehand
within a coherent mathematical framework~\cite{Bogner2010}.
In such a case momentum-dependent \textit{NN} potentials 
are calculated to reproduce exactly the on-shell
amplitudes within a predefined momentum interval.
Although in principle the scheme we have discussed here can also be
extended to $A\!=\!1$, corresponding to \textit{NN} scattering,
we leave this interesting case for a more focused study.

\section{Summary and conclusions}
\label{sec:summary}
  We have investigated the role of high momentum components 
  of microscopic optical model potentials for nucleon-nucleus
  scattering by studying its incidence on the nonlocal
  structure in coordinate space.
  The study considers closed-shell nuclei with mass number in the range 
  $4\!\leq\! A\!\leq\! 208$, for energies from tens of MeV up to 1~GeV.
  To this purpose
  microscopic optical model potentials were constructed in momentum space
  using Bruckner-Hartree-Fock $g$ matrices based on AV18 and N3LO
  chiral potentials.
  We confirm that the gradual suppression of high-momentum contributions 
  of the optical potential results in quite different 
  coordinate-space counterparts, all of them accounting for the
  same scattering observables within a specified accuracy.
  Furthermore,
  we obtain a minimum cutoff momentum $Q$, a function of the target mass
  number and energy of the process, that filters out irrelevant 
  ultraviolet components of the potential.
  We have also found 
  that ultraviolet suppression to PB-type nonlocal potentials
  or local Woods-Saxon potentials results in nonlocal potentials 
  with similar appearance to those based on
  microscopic models in momentum space.

With this study we have shown that, for a given target and energy,
there is a momentum threshold above which features of the potential
become physically meaningless. 
From the prospective of momentum-space optical potential
calculations, such as those investigated in 
Refs.~\cite{Arellano1990,Elster1990,Crespo1990,Vorabbi2016,Aguayo2008,Weppner1998,Arellano2011b},
the identification of $Q$ is particularly useful as it
allows to set reliable 
bounds for the momentum domain over which the potential
needs to be evaluated.
The resulting potentials, referred as \textit{irreducible}
in Ref.~\cite{Arellano2018}, appear to have similar structure in
coordinate space. 

Optical potentials in coordinate space can be expressed in local, 
nonlocal or hybrid forms. Interestingly, we have found that when
these potentials get suppressed their ultraviolet components above
the threshold momentum $Q$, they all share comparable nonlocal features.
Conversely, manifest differences among local, nonlocal or hybrid potentials
rely on the inclusion of Fourier components irrelevant for the
scattering process.
Consequently, it is safe to state that 
a true comparison of nonlocal features of alternative potentials
for a given scattering process require the suppression of 
their ultraviolet components, otherwise the comparison becomes 
with limited scope.

\appendix
\numberwithin{equation}{section}

\section{Multipoles of Gaussian form factor}
\label{awwll}

We evaluate
\begin{equation}
  \label{eq:wl}
  w_l(b)={b}\int_{-1}^{1} P_l(u)e^{bu}\,du\;,
\end{equation}
with $l$ positive integer. 
For low $l\leq 3$ the evaluation of this integral is direct.
For higher values they become tedious but straightforward.
In such cases we use symbolic manipulation software to evaluate 
explicitly the cases $l\leq 5$, obtaining
\begin{subequations}
  \begin{align}
      w_0(b)/2 =& 
    \sinh b \;; \\
   -{b}\, w_1(b)/2 =& 
    \sinh b + b\cosh\,b \;; \\
    {b^2} w_2(b)/2 =& 
    (3+b^2)\sinh\,b - 3b\cosh\,b \;; \\
   -b^3 w_3(b)/2 =& 
    (15+b^2)\sinh b - \nonumber \\ &(15\,b+2\,b^3)\cosh b \;; \\
    b^4 w_4(b)/2 =& 
    (105+45\, b^2+b^4)\sinh b - \nonumber\\& (105\,b+10\,b^3)\cosh b \;; \\
   -b^5 w_5(b)/2 =& 
    (945+420\,b^2+b^4)\sinh b - \nonumber\\&(945\,b+105\,b^3+b^5)\cosh b\;. 
  \end{align}
\end{subequations}
Factorization by exponentials result in
\begin{subequations}
  \begin{align}
    w_0(b) =& 
    e^{b} - e^{-b}\;\\
    w_1(b) =& 
    e^{b}  \left (1 - \frac{1}{b} \right )
  - e^{-b} \left (1 + \frac{1}{b} \right ) \;\\
    w_2(b) =& 
    e^{b}  \left (1 - \frac{3}{b} + \frac{3}{b^2} \right )
  - e^{-b} \left (1 + \frac{3}{b} + \frac{3}{b^2} \right ) \;\\
    w_3(b) =& 
    e^{b}  
    \left (
    1 - \frac{6}{b} + \frac{15}{b^2} - \frac{15}{b^3} 
    \right ) - \nonumber \\
  & e^{-b} 
    \left (
    1 + \frac{6}{b} + \frac{15}{b^2} + \frac{15}{b^3} 
    \right ) \;\\
    w_4(b) =& 
    e^{b}  
   \left (
    1 - \frac{10}{b} + \frac{45}{b^2} - \frac{105}{b^3} + \frac{105}{b^4} 
    \right ) - \nonumber \\
  & e^{-b} 
    \left (
    1 + \frac{10}{b} + \frac{45}{b^2} + \frac{105}{b^3} + \frac{105}{b^4} 
    \right ) \;\\
    w_5(b) =& 
  e^{b}  
   \left (
    1 - \frac{15}{b} + \frac{105}{b^2} - \frac{420}{b^3} + \frac{945}{b^4} 
      - \frac{945}{b^5}
    \right ) - \nonumber \\
 &e^{-b} 
    \left (
    1 + \frac{15}{b} + \frac{105}{b^2} + \frac{420}{b^3} + \frac{945}{b^4} 
      + \frac{945}{b^5}
    \right ). 
  \end{align}
\end{subequations}
Here we recognize Bessel polynomials ${y}_n(x)$ given by
\begin{subequations}
\begin{align}
  {y}_0(x) &= 1 \\
  {y}_1(x) &= x+1 \\
  {y}_2(x) &= 3x^2+3x+1 \\
  {y}_3(x) &= 15x^3+15x^2+6x+1 \\
  {y}_4(x) &= 105x^4+105x^3+45x^2+10x+1 \\
  {y}_5(x) &= 945x^5+945x^4+420x^3+105x^2+15x+1 
\end{align}
\end{subequations}
Thus, 
\begin{equation}
w_l(b)= e^{b}{y}_l \left(\textstyle{\frac{-1\,}{b}} \right ) 
       -e^{-b}{y}_l \left(\textstyle{\frac{\,1\,}{\,b\,}} \right )\;.
\end{equation}
We note that Bessel polynomials are related to 
modified Bessel functions of the second kind through
\begin{equation}
  \label{secondkind}
  y_n(x) = \sqrt{\frac{2}{\pi x}} e^{1/x} K_{n+\nicefrac{1}{2}}(1/x)\;.
\end{equation}
Furthermore, they satisfy the recursion relation
\begin{equation}
  \label{recursion}
  y_{n+1}(x) = (2n+3)x\,y_n(x) + y_{n-1}(x)\;.
\end{equation}
%
%
%
%
%
%
%   BibTeX users please use
%  \bibliographystyle{elsarticle-num}
%  \bibliography{misreferencias}
%  \end{document}

\end{document}